# Orientation Reading by Production Vision-Language Models on Optotype Charts: A Controlled Multi-Model Evaluation Across Reasoning Modes, Prompts, and Access Modalities

**Authors:** Shahryar Wasif[1], Avneek Sandhu[1], and Bin Hu[1,2]

**Affiliations:**

1. Canadian Open Digital Health (OpenDH) Program, University of Calgary, Calgary, Alberta, Canada T2N 4N1
2. Division of Translational Neuroscience, Department of Clinical Neurosciences, Hotchkiss Brain Institute, University of Calgary, Calgary, Alberta, Canada T2N 4N1

**Correspondence should be addressed to:**
**Professor Bin Hu, MD, Ph.D.**
**Suter Professor for Parkinson's Disease Research**
**Founder and Director**
**Open Digital Health (OpenDH) Program, University of Calgary**
**Email: hub@ucalgary.ca**

Abstract:

OBJECTIVES: Vision-language models are increasingly used to interpret medical and everyday images through consumer chat interfaces, yet their ability to read orientation - the single perceptual operation tested by the tumbling-E acuity optotype - is poorly characterized on the surfaces through which they are actually used.

METHODS: We evaluated four production vision-language models (referred to as Claude, GPT, GROK, and Gemini) through their consumer chat interfaces on a locked set of seven optotype charts: four uniform tumbling-E charts (one per cardinal orientation), two mixed-orientation tumbling-E charts, and one Snellen letter chart as a specificity control. Each model was run in two reasoning modes (Fast and Thinking) under two prompt variants (with and without an explicit orientation-decoding rule) by up to three operators. The corpus comprised 920 scoreable trials and 50,420 glyph judgements. The primary outcome was glyph-level accuracy against the chart's designed orientation, summarized with Wilson 95% confidence intervals.

RESULTS: Accuracy ranged from 43.0% to 97.0% across models on identical charts, and the strongest model depended on reasoning mode (GPT 97.0% in Fast mode; GROK 96.6% in Thinking mode). Errors were not random but collapsed onto a model-specific attractor direction. Models were 96-100% internally self-consistent yet ranged widely in accuracy, dissociating reliability from validity. An answer-key-free ensemble-consensus estimate tracked accuracy closely (r = 0.998). For one model, consumer-interface accuracy fell 25-27 points below programmatic access, almost entirely on a single orientation.

CONCLUSIONS: A single accuracy figure conceals clinically relevant, orientation-specific failure modes; vision-language models should be evaluated along multiple axes and on the deployment surface before image-interpretation outputs are trusted.

## Introduction

Standardized optotype charts, on which a patient identifies symbols of decreasing size, are the basis of clinical visual-acuity measurement; for patients who cannot read Roman letters - young children, those who are illiterate, or readers of other scripts - the tumbling-E chart presents a single letter E rotated into one of four cardinal directions and asks the patient to report its orientation.[1] Because every glyph is the same shape and differs only in orientation and size, the tumbling-E chart isolates two perceptual operations, resolving fine spatial detail and determining orientation, that lettered charts entangle with alphabetic recognition.[1]

Large language models have rapidly acquired the capacity to interpret images as well as text, and these multimodal "vision-language" models are now being evaluated across medicine, from answering medical examination questions to interpreting clinical images, including retinal photographs.[2,3,4] Increasingly, patients and clinicians reach these systems not through controlled programmatic interfaces but through the same consumer chat applications used for everyday queries, where a single uploaded photograph can elicit an unsupervised interpretation.[5] How these models behave on simple, well-characterised visual tasks therefore bears directly on whether their image interpretations can be trusted at the point of use.

Despite their fluency, vision-language models have a documented weakness in spatial and geometric reasoning. Production systems pair a Vision-Transformer image encoder, which divides an image into fixed patches and embeds them as tokens,[6] with a contrastive image-text training objective of the CLIP family that optimises semantic alignment rather than geometric fidelity.[7] On benchmarks that isolate perception from semantics, leading models perform close to chance, with spatial-relation and multi-view tasks among the most difficult, and they often behave as though the order of words and objects carries little information.[8,9] These failures have been linked to how image-directed attention is allocated within the model[10] and to a long-recognised bias of visual systems toward cardinal (vertical and horizontal) orientations, the "oblique effect."[11] The tumbling-E chart, whose entire information content is orientation, is thus an unusually clean probe of precisely the capability these models are weakest at.

Most published evaluations report a single aggregate accuracy obtained under one access path and one prompt, which can obscure where and why a model fails and need not reflect performance on the consumer interfaces through which the models are actually used. We therefore evaluated four production vision-language models on a locked set of seven optotype charts, under two reasoning modes and two prompt variants, across multiple independent operators, accessed through their consumer chat interfaces. Our aims were to quantify how much the choice of model and the choice of reasoning mode each move accuracy on identical stimuli; to localise failures by stimulus orientation and characterise the directional structure of errors; to test whether answer-key-free signals - internal self-consistency and inter-model consensus - can estimate accuracy; and to compare consumer-interface performance against programmatic access for one model family.

## Methods

### Experimental design

This was a controlled, observational evaluation with a factorial structure crossing four factors - model (four levels), reasoning mode (two levels), prompt variant (two levels), and operator (up to three) - applied to a fixed, locked set of seven optotype charts. The study evaluated deployed commercial systems on fixed stimuli and involved no human participants or patient data; it was therefore an exploratory measurement study rather than a pre-registered confirmatory trial, and the analysis reports effect sizes and interval estimates rather than hypothesis tests.

### Stimuli

Seven charts were prepared and locked before data collection. Four were uniform tumbling-E charts in which every glyph pointed the same way, one chart each for the LEFT, RIGHT, UP, and DOWN orientations; two were mixed-orientation tumbling-E charts in which glyph orientation varied within the chart; and one was an out-of-family Snellen letter chart, included as a specificity control to test whether a model would force orientation labels onto conventional letters. Each tumbling-E chart contained 61 glyphs arranged in eleven rows of increasing density, following the design principle that letter size is the only quantity that varies systematically down the chart.[1] The designed orientation of every glyph, read directly from the locked stimulus image, constituted the ground truth for scoring.

### Models, access, and reasoning modes

Four production vision-language models, referred to throughout as Claude, GPT, GROK, and Gemini, were accessed through their consumer chat interfaces. Each model was run in two vendor-provided reasoning modes: a default "Fast" mode and an extended-reasoning "Thinking" mode.

Because the reasoning-mode toggle is a product control whose underlying compute policy differs across vendors, mode comparisons are between deployed settings rather than normalised compute budgets. For the Claude family, glyph-level accuracy obtained on the same stimuli through programmatic (API) access was carried as an additional access-modality condition under the full and reduced prompts; these figures were treated as a descriptive reference, not as a target to be matched.

**Prompts**

Two prompt variants were used. The full structured prompt contained an explicit spine-based decoding rule mapping glyph geometry to a reported direction; the reduced prompt removed that rule while preserving task framing and output format, so that the two variants differed in a single component. Comparing accuracy under the two prompts (full minus reduced, in percentage points) isolated the contribution of the decoding rule for each model and mode.

**Procedure and operators**

Trials were run by up to three independent operators. Not every operator ran every model; the matched four-model agreement analysis therefore rests on the sessions of the single operator who ran all four models, and inter-operator analyses are limited accordingly, as noted where they apply. For each trial, an operator submitted a chart image to a model under a specified mode and prompt and recorded the model's per-glyph orientation responses and the wall-clock trial duration. The corpus comprised 920 scoreable trials and 50,420 individual glyph judgements.

**Scoring and ground truth**

Each glyph judgement was scored against the chart's designed orientation. Glyph omissions and additions were tracked separately from direction errors throughout scoring, so that a mis-oriented glyph and a mis-counted glyph were treated as distinct events. Scoring was independently re-verified on a 30-trial random sample plus six targeted spot-checks, with zero discrepancies.

**Outcomes**

The primary outcome was glyph-level accuracy: the proportion of per-chart glyph judgements whose reported direction matched the designed orientation, pooled across prompts and operators within each model × mode cell. Pre-specified secondary outcomes were accuracy stratified by stimulus orientation; the directional distribution of errors; the prompt effect (full minus reduced); internal self-consistency (the proportion of uniform-chart trials on which every glyph received the same label, a reliability measure computable without the answer key); inter-model agreement and agreement with an answer-key-free ensemble consensus (the majority label across the pool of models, modes, and operators); wall-clock trial latency; prompt sensitivity and inter-operator reproducibility of the modal committed direction; out-of-family (Snellen) flag rate; and structural format fidelity (the proportion of trials reporting exactly the 61 glyphs present by design).

**Statistical analysis**

Glyph-level accuracy is reported with Wilson score 95% binomial confidence intervals, which have better small-sample coverage than the normal-approximation interval and are bounded within the unit interval.[12],[13] Inter-model agreement was quantified by pairwise percent agreement and by Fleiss' κ on the matched glyph positions available from the operator who ran all four models;[14] κ values were interpreted using the Landis-Koch descriptive bands.[15] The correspondence between agreement with the ensemble consensus and measured accuracy across the eight model × mode cells was summarised with the Pearson correlation coefficient. Wall-clock durations were recorded by operators with a stopwatch in heterogeneous formats, parsed to seconds, restricted to a 1-600-second plausibility window, and anchored on the operator whose formatting was internally consistent across all eight cells, with the ordering confirmed against the other operators; latency is therefore reported as an exploratory efficiency signal that includes operator handling and output generation rather than isolated inference time.

## Results

We evaluated four production vision-language models accessed through their consumer chat interfaces - referred to throughout as Claude, GPT, GROK, and Gemini - on a locked set of seven optotype charts. Four charts were uniform tumbling-E charts in which every glyph points the same way (one chart each for the LEFT, RIGHT, UP and DOWN orientations); two were mixed-orientation tumbling-E charts; and one was an out-of-family Snellen letter chart used as a specificity control. Each tumbling-E chart contained 61 glyphs in eleven rows of increasing density. Every model was run in two reasoning modes (a default "Fast" mode and an extended-reasoning "Thinking" mode), under two prompt variants (a full structured prompt containing an explicit spine-based orientation-decoding rule, and a reduced prompt with that rule removed), by up to three independent operators. The corpus comprised 920 scoreable trials and 50,420 individual glyph judgements. Each glyph judgement was scored against the chart's designed orientation, which constitutes the ground truth for this task and was read directly from the locked stimulus images; scoring was independently re-verified on a 30-trial random sample plus six targeted spot-checks with zero discrepancies. For one model family (Claude), figures from a prior programmatic-access (API) evaluation of the same stimuli are carried as an additional access-modality condition, not as a target to be matched.

Results are organised as follows. We first report the primary accuracy outcome by model and reasoning mode (3.1), then decompose it by stimulus orientation to localise failures (3.2) and by the directional structure of errors (3.3). We then report the effect of the prompt variant (3.4), the relationship between internal reliability and accuracy (3.5), inter-model agreement and a consensus-based quality estimate (3.6), the accuracy-latency trade-off of reasoning mode (3.7), prompt sensitivity and inter-operator reproducibility (3.8), out-of-family discrimination and format fidelity (3.9), and finally the API-versus-UI access-modality comparison (3.10).

### 3.1 Accuracy by model and reasoning mode

*Analysis and rationale.* The primary outcome is glyph-level accuracy: the proportion of the 61 per-chart glyph judgements whose reported direction matched the chart's designed orientation, pooled across prompts and operators within each model×mode cell. We report Wilson 95% binomial confidence intervals. The goal is to quantify, on identical stimuli and identical user-level instructions, how much the choice of model and the choice of reasoning mode each move accuracy.

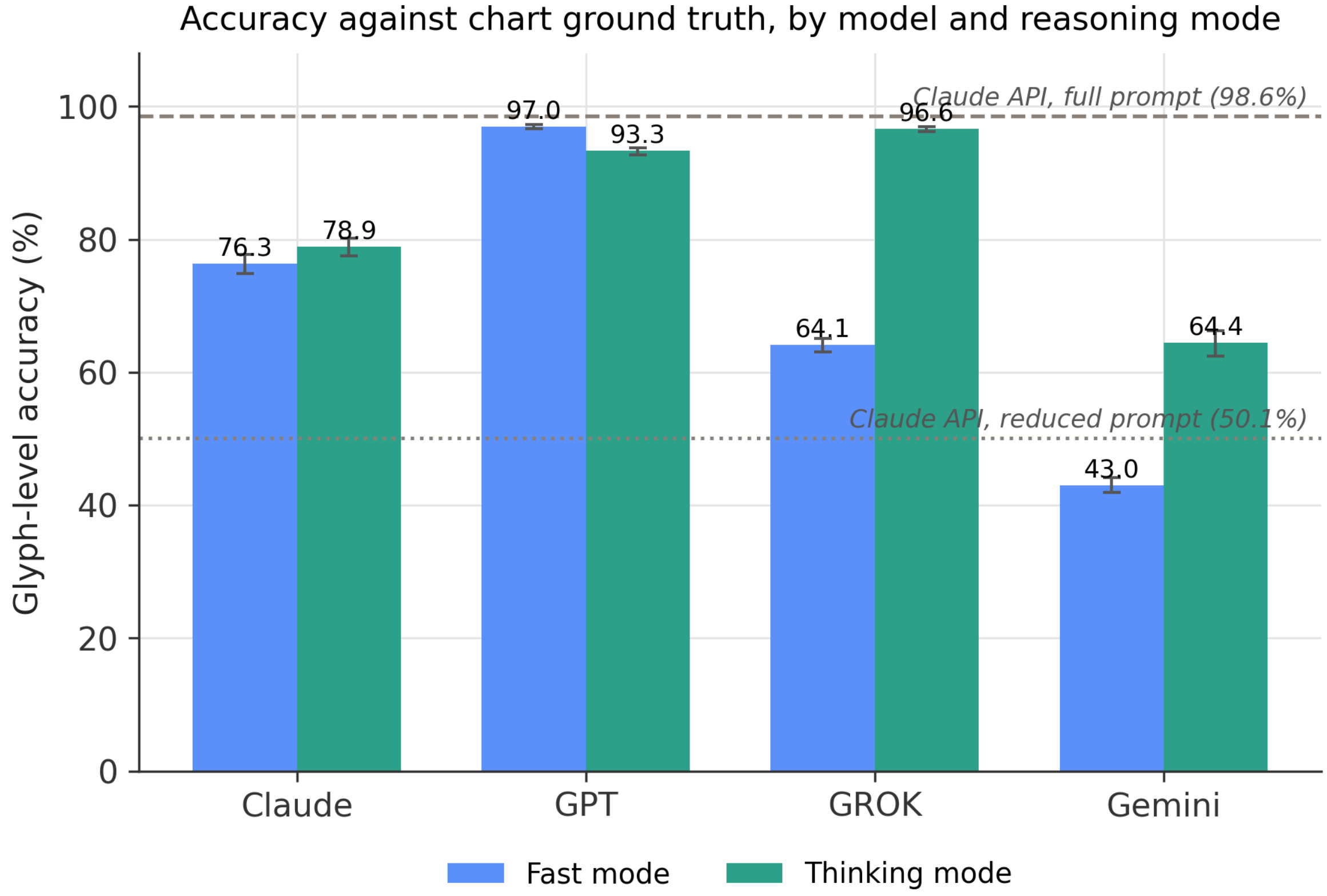

***Figure 1.*** *Glyph-level accuracy against chart ground truth by model and reasoning mode, pooled across both prompt variants and all operators. Error bars are Wilson 95% binomial confidence intervals. Horizontal reference lines mark the Claude API figures for the full prompt (98.6%) and reduced prompt (50.1%), included as an access-modality reference rather than as a target.*

*Findings.* Accuracy varied widely across models on identical charts (Figure 1). In Fast mode, GPT was most accurate at 97.0% (95% CI 96.6-97.3), followed by Claude at 76.3% (74.9-77.7), GROK at 64.1% (63.1-65.1), and Gemini at 43.0% (41.9-44.1) - a 54.0-percentage-point spread between the strongest and weakest model. In Thinking mode the ordering changed: GROK rose to 96.6% (96.2-97.0) and GPT fell slightly to 93.3% (92.7-93.8), so that GROK and GPT were the two strongest Thinking-mode readers; Claude was 78.9% (77.5-80.2) and Gemini 64.4% (62.4-66.3). The reasoning-mode change was therefore large and model-specific: GROK gained 32.5 points and Gemini 21.4 points from Thinking mode, Claude changed by only +2.6 points, and GPT declined by 3.7 points.

### 3.2 Accuracy by stimulus orientation

*Analysis and rationale.* To localise where accuracy is gained or lost, we stratified accuracy by the chart's designed orientation. Because the four uniform charts are a balanced set (one per direction), this stratification isolates whether a model handles each of the four orientations equally or fails selectively on particular ones. This directly tests whether aggregate accuracy conceals orientation-specific blind spots.

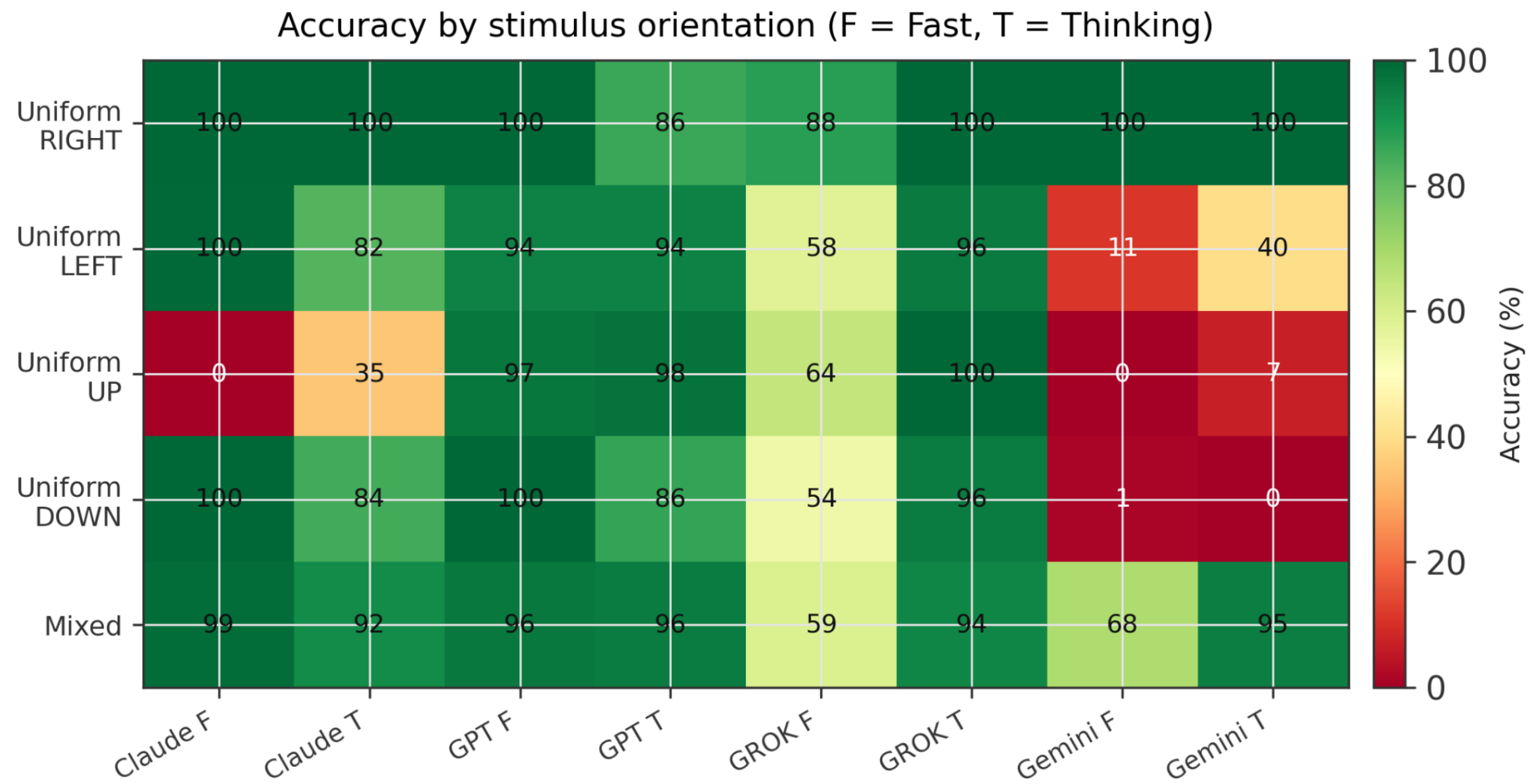

***Figure 2.*** *Glyph-level accuracy by stimulus orientation for each model x mode cell (F = Fast, T = Thinking). Cells are coloured from red (0%) to green (100%). The out-of-family Snellen control is analysed separately in Section 3.9.*

*Findings.* Aggregate accuracy concealed sharp orientation-specific structure (Figure 2). Every model read the canonical uniform-RIGHT chart almost perfectly (88.1-100%). The non-RIGHT orientations separated the models. The uniform-UP chart was the single most discriminating stimulus: GPT (96.6% Fast, 97.5% Thinking) and GROK-Thinking (100%) read it accurately, whereas Claude scored 0.0% in Fast mode and 35.1% in Thinking mode, and Gemini scored 0.0% (Fast) and 6.8% (Thinking). Claude's UP failure is notable because Claude simultaneously scored 100% on the uniform-DOWN chart in Fast mode and 99.5% on uniform-LEFT - the deficit was specific to a single orientation. Gemini failed three of four non-RIGHT orientations in Fast mode (LEFT 11.0%, UP 0.0%, DOWN 1.4%). GROK's Fast-mode weakness was distributed across LEFT (58.1%), UP (64.3%) and DOWN (54.1%) and was largely repaired by Thinking mode (95.6-100%). Mixed charts were handled well by the stronger readers (GPT 96.3% Fast; Claude 98.7% Fast) and poorly by Gemini Fast (68.1%).

### 3.3 Directional structure of errors

*Analysis and rationale.* To characterise the nature of incorrect responses rather than merely their frequency, we classified every individual glyph error by the direction the model reported in place of the correct one. If errors were random, the misreported directions would be roughly uniform; a concentration on one direction indicates a systematic attractor. We computed this distribution per model×mode cell over all glyph errors (n = 10,416 errors in total).

*Findings.* Errors were far from random; each model×mode cell had a dominant misreport direction (Supplementary Figure S1). GROK in Fast mode directed 73% of its 2,956 errors to RIGHT. Gemini directed the majority of its errors to RIGHT in both modes (59% of 4,700 Fast-mode errors; 67% of 804 Thinking-mode errors). Claude's errors concentrated instead on DOWN (80% of 811 Fast-mode errors; 53% of 736 Thinking-mode errors), the great majority arising from the uniform-UP chart being read as DOWN. GPT, which made few errors in Fast mode, showed a mixed RIGHT/DOWN pattern there (53%/35% of 242 errors), but in Thinking mode its 549 errors concentrated on UP (80%), reflecting the Thinking-mode decline in uniform-RIGHT accuracy noted in Section 3.2. Across cells, a single direction accounted for between 53% and 80% of all errors, confirming that the dominant failure mode is a directional attractor rather than diffuse noise; the specific attractor was RIGHT for GROK and Gemini, DOWN for Claude, and UP for GPT-Thinking.

### 3.4 Effect of the prompt variant

*Analysis and rationale.* The two prompt variants differed in a single component: the full prompt contained an explicit spine-based decoding rule mapping glyph geometry to a direction, while the reduced prompt removed it but preserved task framing and output format. Comparing accuracy under the two prompts (full minus reduced, in percentage points) measures whether this instructional component helped, was neutral, or hurt, for each model×mode cell. We treat this as a factor in its own right rather than as a contrast to be matched against any external benchmark.

*Findings.* The prompt effect was strongly model-dependent and not uniformly positive (Supplementary Figure S2). For GROK, the full prompt helped substantially in Fast mode (76.8% vs 51.9%; +24.9 pp) and modestly in Thinking mode (98.5% vs 94.7%; +3.8 pp). For GPT the prompt was essentially neutral in both modes (+0.4 pp Fast, +1.8 pp Thinking), consistent with GPT already reading the charts near-ceiling without the rule. For Claude and Gemini the full prompt was associated with lower accuracy than the reduced prompt: Claude -8.3 pp (Fast) and -9.4 pp (Thinking), Gemini -7.6 pp (Fast) and -12.9 pp (Thinking). The decoding rule therefore behaved as an active and beneficial component only for GROK; for two of the four models its presence coincided with a reliable accuracy decrement.

### 3.5 Internal reliability and its relationship to accuracy

*Analysis and rationale.* Because every glyph on a uniform chart is identical, a model that reads reliably should report one repeated direction for the whole chart, irrespective of whether that direction is correct. We therefore measured internal self-consistency as the proportion of uniform-chart trials in which every glyph received the same label. This is a reliability measure computable without the answer key. Placing it beside accuracy (a validity measure) tests whether the two coincide or dissociate - an important question because a consistency check is a natural first screen when ground truth is scarce.

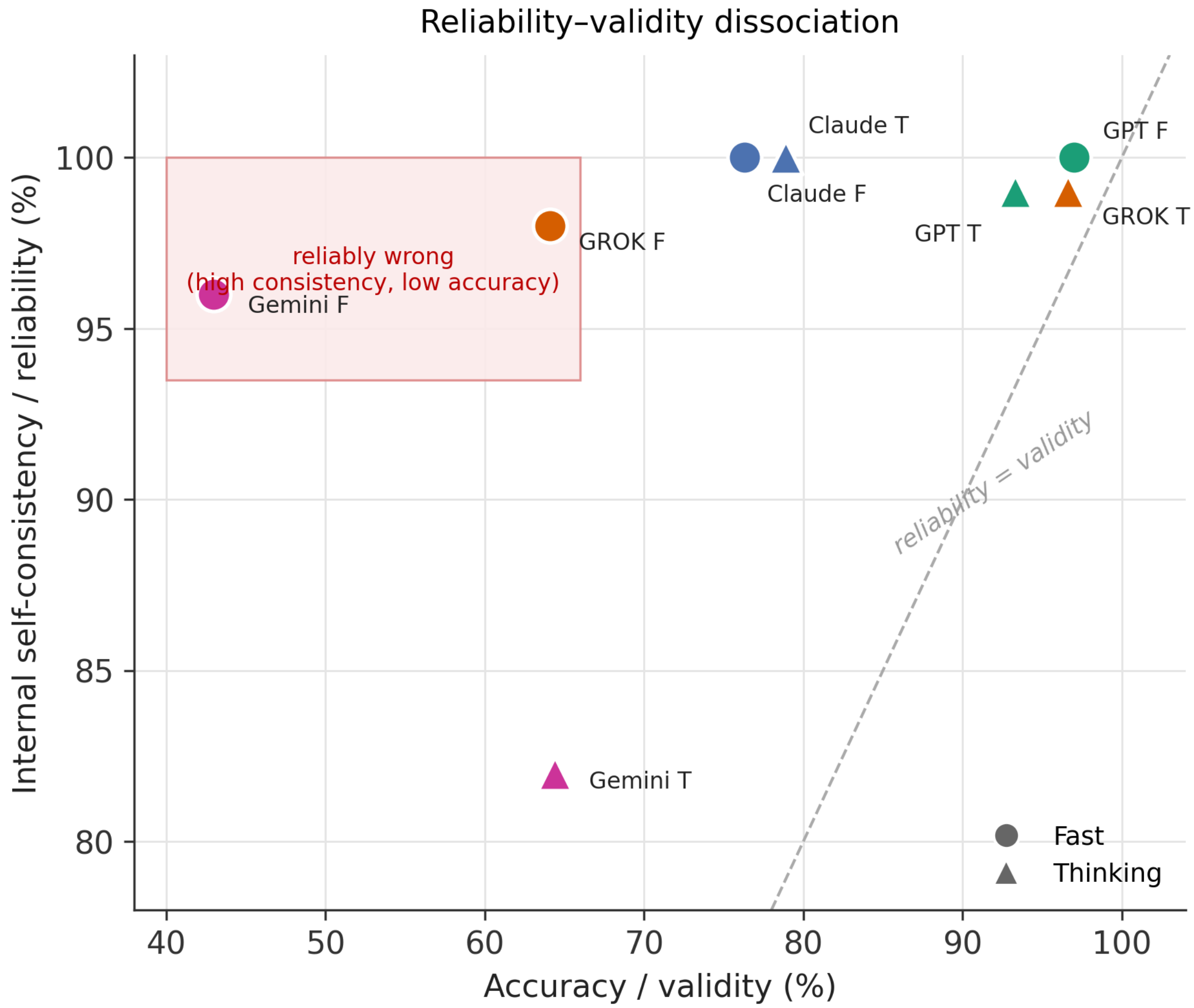

***Figure 3.*** *Reliability-validity dissociation. Each point is a model x mode cell positioned by accuracy (x-axis) and internal self-consistency (y-axis). Points far above the diagonal are internally consistent but inaccurate ("reliably wrong"); points on the diagonal are equally reliable and valid. Circles = Fast; triangles = Thinking.*

*Findings.* All four models were highly internally consistent: the proportion of fully internally-uniform trials was 96-100% in Fast mode (Claude 100%, GPT 100%, GROK 98%, Gemini 96%) and 82-99% in Thinking mode. Consistency and accuracy nonetheless dissociated sharply (Figure 3). Gemini in Fast mode was 96% internally consistent yet only 43.0% accurate - a 53-point reliability-validity gap, the signature of a model that confidently and repeatedly reports the same incorrect direction.

GROK in Fast mode showed a 34-point gap (98% consistent, 64.1% accurate) and Claude a 21-24-point gap. GPT and GROK-Thinking closed the gap to 2-6 points, being both reliable and accurate. The practical consequence is that internal-consistency screening alone would have certified the least accurate cell (Gemini Fast) as well-behaved; reliability is necessary but not sufficient for validity on this task.

### 3.6 Inter-model agreement and a consensus-based quality estimate

*Analysis and rationale.* Treating the four models as independent raters of the same glyphs, we quantified how often they agreed, using pairwise percent agreement and Fleiss' κ on the matched glyph positions available from the operator who ran all four models. We then asked whether a purely relative signal - the majority label across the full pool of models, modes and operators - could approximate accuracy. For each model×mode cell we computed its agreement with this ensemble consensus and compared it against measured accuracy. The goal is to test whether model quality can be estimated when an answer key is unavailable.

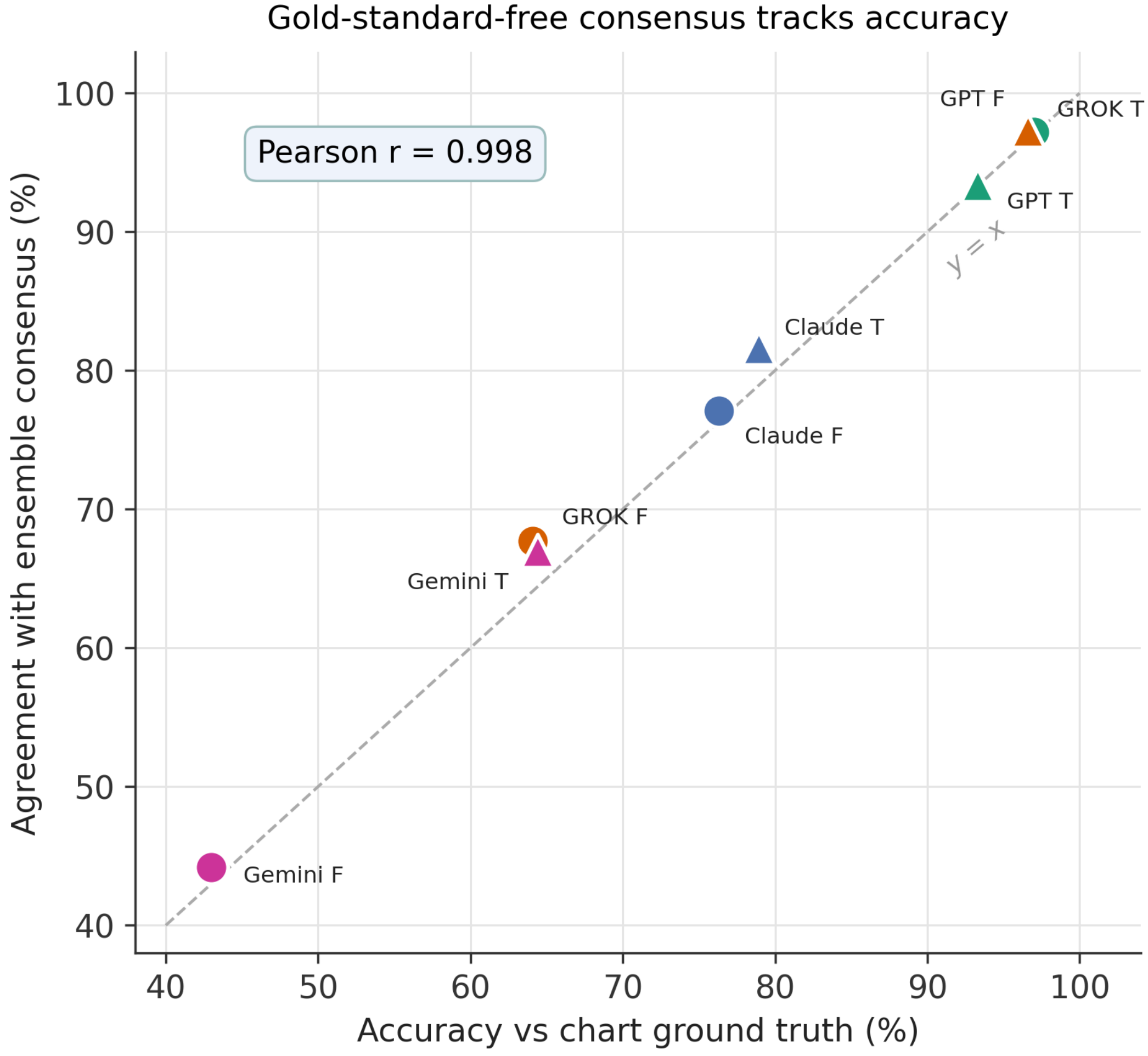

***Figure 4.*** *Agreement with the gold-standard-free ensemble consensus (y-axis) against accuracy versus chart ground truth (x-axis) for all eight model x mode cells. The dashed line is y = x; points fall almost exactly on it (Pearson r = 0.998).*

*Findings.* Inter-model agreement was modest. All four models reported the same direction on only 31.8% of matched glyph positions in Fast mode and 43.8% in Thinking mode; a three-of-four majority existed for 68.9% (Fast) and 82.6% (Thinking). Fleiss' κ was 0.37 (fair, by the Landis-Koch bands) in Fast mode and 0.53 (moderate) in Thinking mode, so extended reasoning materially increased convergence. Agreement with the ensemble consensus tracked measured accuracy almost exactly across all eight cells (Figure 4; Pearson r = 0.998): for example GPT-Fast 97.2% consensus agreement vs 97.0% accuracy, GROK-Fast 67.7% vs 64.1%, and Gemini-Fast 44.2% vs 43.0%. Because each cell contributes to the consensus it is scored against, this correspondence is a slight upper bound; a leave-one-model-out consensus would widen the small residual gaps. The correspondence holds here because, on most charts, the plurality of cells reads correctly so the consensus lands on the correct direction; on the one chart where the models share a rightward prior (uniform-RIGHT) consensus and accuracy would agree for the wrong reason, a caveat we return to in the Discussion.

### 3.7 The accuracy-latency trade-off of reasoning mode

*Analysis and rationale.* Extended reasoning improves accuracy for some models (Section 3.1) but consumes additional inference time. We recorded wall-clock trial duration and combined it with accuracy to characterise the trade-off. Durations were recorded by operators with a stopwatch in heterogeneous formats; we parsed them to seconds, restricted to a 1-600 s plausibility window, anchored the analysis on the operator whose formatting was internally consistent across all eight cells, and confirmed the ordering with the other operators. These are wall-clock trial times including operator handling and output generation, not isolated inference latency.

*Findings.* Latency differed by an order of magnitude across cells and the accuracy gains from Thinking mode came at a steep time cost (Supplementary Figure S3). GROK in Fast mode was by far the quickest (median 8 s, consistent at 8-9 s across all three operators) but capped at 64.1% accuracy; switching GROK to Thinking mode raised accuracy to 96.6% but increased median latency roughly eleven-fold to 85 s. GPT reached 97.0% accuracy in Fast mode at 70 s without needing Thinking mode, while its Thinking mode was both slightly less accurate (93.3%) and slower (104 s). Gemini was fast in both modes (11 s, 46 s) but never accurate (43.0%, 64.4%). The most favourable accuracy-at-acceptable-latency operating points were thus GPT-Fast and GROK-Thinking, reached by opposite routes - GPT by a capable single pass, GROK by spending reasoning time.

### 3.8 Prompt sensitivity and inter-operator reproducibility

*Analysis and rationale.* Two further reliability properties bear on whether a measured accuracy can be trusted to recur. Prompt sensitivity asks whether a model's committed direction on a fixed chart changes between the two prompt variants - a model whose reading flips with cosmetic prompt changes is harder to deploy predictably. Inter-operator reproducibility asks whether two operators running the identical model, mode, prompt and chart obtain the same committed direction. We computed both at the level of the modal committed direction per uniform chart.

*Findings.* Prompt sensitivity varied from negligible to severe. Claude's committed direction was identical under both prompts on every uniform chart in both modes (100% agreement) - it was effectively prompt-insensitive. GPT was largely insensitive (92% Fast, 100% Thinking). GROK in Fast mode changed its committed direction on a majority of charts between the two prompts (only 42% agreement), i.e. it was highly prompt-sensitive, but became fully insensitive in Thinking mode (100%). Gemini was moderately sensitive (75% Fast, 50% Thinking). Inter-operator reproducibility was high for most cells but not all: it was 100% for Claude Fast, GPT Thinking and GROK Thinking, but only 60% for Claude Thinking and 50% for GROK Fast, meaning that for those cells operators sometimes obtained different committed directions under nominally identical conditions. Reproducibility was generally higher in Thinking mode. These figures bound how much cross-condition variation could be operator-induced and argue for reporting operator counts in any manual-interface evaluation.

### 3.9 Out-of-family discrimination and format fidelity

*Analysis and rationale.* Two behavioural measures probe whether a model tracks the stimulus rather than pattern-matching to the expected task. Out-of-family discrimination is the rate at which a model flags the Snellen letter chart as not belonging to the tumbling-E family rather than forcing orientation labels onto conventional letters; a model that scores the Snellen chart has mis-identified the stimulus class. Format fidelity is the rate at which a model reports exactly the 61 glyphs present by design; over- or under-counting indicates the model is not faithfully tracking chart structure.

*Findings.* Discrimination was strongly mode-gated for two models. Claude flagged the Snellen chart in 100% of trials in both modes. GPT and GROK flagged it reliably only in Thinking mode (GPT 64% Fast → 100% Thinking; GROK 53% Fast → 100% Thinking). Gemini did not reliably flag it in either mode (73% Fast, 60% Thinking). Format fidelity also separated the models: GPT in Thinking mode reported exactly 61 glyphs on 96% of trials, whereas GROK matched the count on only 16% (Fast) and 26% (Thinking) of trials, with a median report of 66 glyphs, indicating systematic over-segmentation. Claude and Gemini were intermediate (55-76%). Over-counting is a distinct failure mode from mis-orienting a correctly counted glyph and complicates alignment of reported labels to chart positions; glyph omissions and additions were tracked separately from direction errors throughout scoring.

### 3.10 Access modality: API versus consumer interface

*Analysis and rationale.* For the Claude family, glyph-level accuracy on these same stimuli was available from a prior programmatic (API) evaluation under both prompt variants. We carry those figures as an additional access-modality condition - API versus consumer chat interface (UI) - holding the model family, task and prompt content constant. The comparison is descriptive: the two access paths differ in image pre-processing, system-prompt layering, tool affordances, sampling defaults and version routing, none of which we can separate here, so the analysis localises that an access-modality difference exists and where it concentrates, not which sub-mechanism produces it. The API figures are not treated as a standard to be matched.

*Findings.* Under the full prompt, Claude via API achieved 98.6% overall accuracy, whereas Claude via the UI achieved 71.7% in Fast mode and 73.8% in Thinking mode - a difference of roughly 25-27 percentage points at the aggregate level. Decomposed by orientation (Supplementary Figure

S4), the gap was overwhelmingly localised to a single orientation: API Claude read the uniform-UP chart at 100% while UI Claude read it at approximately 0% in Fast mode, with the other orientations broadly comparable across the two access paths (RIGHT, LEFT and DOWN all near-ceiling in both; mixed charts high in both). In other words, the access-modality difference for Claude on this task was not a uniform degradation but an orientation-specific collapse confined to the UP chart in the UI condition. The aggregate API-UI gap is therefore best read as the average of a near-complete UP-orientation failure and otherwise comparable performance, rather than a broad across-the-board deficit.

## Discussion

### Principal findings

On an identical, tightly controlled optotype-orientation task, four production vision-language models differed by up to 54 percentage points in glyph-level accuracy under identical user instructions, and their relative ordering depended on reasoning mode: GPT was strongest in Fast mode (97.0%) while GROK was strongest in Thinking mode (96.6%), and Gemini was weakest throughout (43.0-64.4%). Aggregate accuracy concealed sharp orientation-specific structure: the uniform-UP chart was read near-perfectly by GPT and GROK-Thinking but at essentially 0% by Claude-Fast and Gemini-Fast, and the errors were not random but collapsed onto a model-specific attractor direction (RIGHT for GROK and Gemini, DOWN for Claude, UP for GPT-Thinking). The four models were uniformly highly self-consistent (96-100% in Fast mode) yet this reliability dissociated from validity - Gemini-Fast was 96% self-consistent but 43% accurate - demonstrating that internal consistency cannot substitute for accuracy. An ensemble-consensus quality estimate computed without any answer key tracked accuracy almost exactly ($r = 0.998$). Extended reasoning increased inter-model agreement (Fleiss' $\kappa$ 0.37→0.53) and rescued accuracy for GROK and Gemini, but at up to an eleven-fold latency cost. The decoding-rule prompt component helped only GROK and was associated with reduced accuracy for Claude and Gemini. We now consider the mechanisms that plausibly account for these patterns.

**Why orientation is hard, and why a canonical direction dominates errors**

The error structure we observed - high accuracy on the canonical RIGHT-facing E and collapse of errors onto a single dominant direction - is consistent with how contemporary vision-language models encode images and with the statistics of their training data. Production VLMs pair a vision encoder, typically a Vision Transformer that splits the image into fixed patches and embeds them as tokens,[6] with a contrastive image-text pre-training objective of the CLIP family.[7] That objective optimises for semantic image-text alignment rather than geometric fidelity,[7] and a growing literature documents the resulting weakness in spatial and orientation reasoning. On benchmarks that isolate perception from semantics, leading models perform close to chance, with spatial-relation and multi-view (rotation-like) tasks among the hardest categories.[8] A complementary line of work shows that these models frequently behave as though the order and configuration of image and text elements carry little information - a "bag-of-words" failure mode that is exactly the kind of order- and orientation-insensitivity an optotype task exposes.[9] Interpretability analyses further localise such errors to the model misdirecting its image-directed attention away from the relevant regions, with attention allocated differently to image than to text tokens.[10] Because the encoder is optimised for semantic rather than geometric correspondence,[7],[9] fine orientation information is represented only weakly, and its recovery is expected to be poorest once a stimulus departs from a canonical, cardinally-aligned pose - the regime in which visual systems are themselves least reliable.[11]

The specific concentration of errors on a canonical orientation has an additional, well-established root in training-data statistics. Natural images are dominated by cardinal (vertical and horizontal) structure, and both biological visual systems and convolutional networks trained on natural images over-represent cardinal orientations - the "oblique effect," explained by efficient coding of the most probable orientations in the visual environment.[11] A model whose visual prior is shaped by predominantly upright, canonically-posed training images will, when its orientation read-out is uncertain, fall back on the most probable orientation. For a letter E that most probable form is the standard right-facing glyph, which is exactly the attractor we observed for GROK and Gemini. Claude's deviation - a DOWN rather than RIGHT attractor, arising specifically from the W/Ш-shaped UP chart being read as the M/m-shaped DOWN chart - is consistent with a confusion between the two vertically-oriented variants that are mirror images about the horizontal axis, a harder discrimination than separating the horizontally-oriented LEFT and RIGHT variants. In all cases the key point is that the dominant error is a systematic prior, not noise; this is why internal consistency remained high even where accuracy collapsed.

**Why reasoning mode helps some models and not others**

Extended-reasoning modes expose additional inference-time computation, the multimodal analogue of chain-of-thought prompting, which improves performance on tasks that benefit from explicit intermediate steps,[16] including multimodal reasoning benchmarks.[17] For an orientation task, the productive intermediate step is to localise the spine of the E and reason explicitly about which way the open side faces, rather than emitting a fast prior-driven guess. This predicts that reasoning mode should help most where the fast prior is wrong and least where the fast pass is already correct - exactly the pattern we observed. GROK and Gemini, whose Fast modes were dominated by a rightward prior, gained 32.5 and 21.4 points respectively; GPT, already near-ceiling in Fast mode, gained nothing and in fact lost 3.7 points; and Claude, whose deficit is a specific UP/DOWN confusion rather than a global prior, was only marginally improved. The accompanying rise in inter-model agreement (κ 0.37→0.53) and the flattening of GROK's rightward skew to a near-uniform directional distribution both indicate that reasoning mode pulls divergent fast priors toward a more geometry-driven, and more concordant, reading. The GPT-Thinking decline, concentrated as an UP attractor on the otherwise-easy RIGHT chart, is consistent with over-deliberation - additional reasoning occasionally overturning a correct fast judgement; we present this as an interpretation of our own data rather than an established result. Finally, the benefit is not free: the eleven-fold latency increase for GROK quantifies the inference-time cost of buying this accuracy, and the existence of GPT-Fast (97% at 70 s with no reasoning step) shows that a sufficiently capable single pass can reach the same accuracy frontier by a different route.

**Reliability without validity, and gold-standard-free quality estimation**

The dissociation between internal consistency and accuracy is the study's central methodological caution. A model that has collapsed onto a directional prior is, by construction, highly self-consistent: it emits the same wrong answer for every identical glyph. Consistency therefore certifies that a model's output is stable, not that it is correct, and a consistency screen - attractive precisely because it needs no answer key - would have passed the least accurate cell in our study. The constructive counterpart to this caution is that a different answer-key-free signal, agreement with an ensemble consensus of independent models, tracked accuracy almost perfectly (r = 0.998). This is the multi-rater analogue of self-consistency decoding, in which agreement across independent reasoning samples is used as a quality signal.[18] Its validity here rests on a condition that must be checked rather than assumed: the consensus approximates truth only when the plurality of raters is correct, so where independent models share a bias - as all four partly do on the canonical-RIGHT chart - the consensus would inherit that bias and agree with accuracy for the wrong reason. The safeguard, which our data support, is to read the consensus signal alongside the directional-preference diagnostic, which flags exactly the shared-prior situations in which consensus would mislead. Used together, these answer-key-free measures provide a practical quality estimate for settings where a definitive orientation key is unavailable, while the reliability-validity dissociation marks the boundary of what consistency alone can establish.

**Access modality and deployment surface**

For the Claude family, the same model on the same stimuli differed by roughly 25-27 aggregate percentage points between API and consumer-interface access, and the difference was almost entirely localised to the uniform-UP orientation rather than distributed across the task. Several surface-level factors differ between the two access paths and could each contribute: consumer interfaces rasterise uploaded images at a server-chosen resolution before the model receives a pixel array, layer a substantial system prompt and tool scaffolding onto the request, and apply sampling and version-routing defaults that the API exposes or fixes. Our design cannot separate these, so we restrict the claim to its defensible form: an access-modality difference exists for this model and task, it is large, and it concentrates on the single orientation that is also the hardest for the model in the interface condition. The orientation-specificity is itself informative - a uniform degradation would point to a global factor such as image down-sampling, whereas a single-orientation collapse points to an interaction between the surface and the model's pre-existing UP/DOWN weakness. The broader implication is that an accuracy figure obtained on one access path does not transfer unchanged to another, and evaluations intended to inform interface-based deployment should be conducted on the deployment surface.

**Implications and limitations**

The results argue for evaluating vision-language models on visual tasks along several axes rather than a single accuracy score. Models that were indistinguishable on one axis diverged on others: GPT and GROK reached the same accuracy ceiling but by opposite latency routes; Claude and GPT were both highly self-consistent but differed by 20 points in accuracy; and every model passed an internal-consistency screen while ranging from 43% to 97% accurate. A compact panel - accuracy where ground truth exists, plus consistency, inter-model concordance, directional-preference structure, prompt sensitivity, reproducibility, discrimination and latency - captured behavioural structure that accuracy alone obscured, and most of these measures are obtainable without an answer key.

Several limitations bound these conclusions. The task is a single, deliberately controlled optotype probe with seven charts and a four-category response space; it was chosen to expose orientation behaviour cleanly and we do not claim that specific numeric values transfer to richer natural-image tasks. Per-cell trial counts were uneven (42-150), and the smaller cells (notably Gemini-Thinking and the per-operator Claude cells) carry wider intervals, reflected in the reported confidence bounds. Not every operator ran every model, so the matched four-model agreement analysis rests on a single operator's sessions and some inter-operator contrasts are limited; this is stated wherever it applies. The models were accessed through consumer interfaces whose image pre-processing, system prompts, sampling and version routing are not observable, and exact model-version strings were not logged at every session, so a portion of cross-cell variation may reflect version differences rather than the factors to which we attribute it. The reasoning-mode toggle is a product control whose underlying compute policy differs across vendors, so mode comparisons are between deployed settings, not normalised compute budgets. Latency was recorded as wall-clock trial duration including operator handling and was parsed from heterogeneous formats under a plausibility window; it is reported as an exploratory efficiency signal rather than precise inference timing. Finally, the analysis is exploratory: we report effect sizes and interval estimates rather than a pre-registered confirmatory test, and the directional and reliability effects, though large relative to their intervals, would benefit from pre-registered replication on additional tasks and stimulus families.

## Conclusions

This controlled evaluation demonstrates that production vision-language models can be highly repeatable yet systematically wrong on a simple orientation task. Performance depended not only on model identity but also on reasoning mode, prompt design, stimulus orientation, and access modality, with errors clustering around model-specific directional attractors rather than diffuse noise. Accordingly, aggregate accuracy and internal self-consistency should not be used alone to judge readiness for clinical or other high-stakes image interpretation. Evaluations should be conducted on the intended deployment surface and should report orientation-stratified accuracy, directional error structure, prompt sensitivity, inter-operator reproducibility, consensus behaviour, 21 and latency before model outputs are trusted. Multi-model consensus may provide a useful answer-key-free quality signal, but only when shared directional bias is assessed separately.

## Acknowledgements

The author acknowledges methodological feedback from large language model reviewers used as thinking partners during protocol development and manuscript drafting. AI methodological assistance is disclosed per emerging community norms; no AI system was a co-author or contributed substantively to the scientific content beyond methodological review and drafting support.

## References

1. Taylor HR. Applying new design principles to the construction of an illiterate E chart. Am J Optom Physiol Opt 1978;55:348-351.

2. Moor M, Banerjee O, Abad ZSH, et al. Foundation models for generalist medical artificial intelligence. Nature 2023;616:259-265.

3. Singhal K, Azizi S, Tu T, et al. Large language models encode clinical knowledge. Nature 2023;620:172-180.

4. Jalili J, Jiravarnsirikul A, Bowd C, et al. Glaucoma detection and feature identification via GPT-4V fundus image analysis. Ophthalmol Sci 2025;5:100667.

5. Ayers JW, Poliak A, Dredze M, et al. Comparing physician and artificial intelligence chatbot responses to patient questions posted to a public social media forum. JAMA Intern Med 2023;183:589-596.

6. Dosovitskiy A, Beyer L, Kolesnikov A, et al. An image is worth 16×16 words: transformers for image recognition at scale. In: Proceedings of the International Conference on Learning Representations (ICLR), 2021.

7. Radford A, Kim JW, Hallacy C, et al. Learning transferable visual models from natural language supervision. In: Proceedings of the 38th International Conference on Machine Learning (ICML). PMLR 2021;139:8748-8763.

8. Fu X, Hu Y, Li B, et al. BLINK: multimodal large language models can see but not perceive. In: Computer Vision - ECCV 2024. Lecture Notes in Computer Science. Cham: Springer, 2024.

9. Yuksekgonul M, Bianchi F, Kalluri P, Jurafsky D, Zou J. When and why vision-language models behave like bags-of-words, and what to do about it? In: Proceedings of the International Conference on Learning Representations (ICLR), 2023.

10. Chen S, Zhu T, Zhou R, et al. Why is spatial reasoning hard for VLMs? An attention mechanism perspective on focus areas. In: Proceedings of the 42nd International Conference on Machine Learning (ICML). PMLR 2025;267:9910-9932.

11. Henderson MM, Serences JT. Biased orientation representations can be explained by experience with non-uniform training set statistics. J Vis 2021;21(8):10.

12. Wilson EB. Probable inference, the law of succession, and statistical inference. J Am Stat Assoc 1927;22:209-212.

13. Brown LD, Cai TT, DasGupta A. Interval estimation for a binomial proportion. Stat Sci 2001;16:101-133.

14. Fleiss JL. Measuring nominal scale agreement among many raters. Psychol Bull 1971;76:378-382.

15. Landis JR, Koch GG. The measurement of observer agreement for categorical data. Biometrics 1977;33:159-174.

16. Wei J, Wang X, Schuurmans D, et al. Chain-of-thought prompting elicits reasoning in large language models. In: Advances in Neural Information Processing Systems (NeurIPS), 2022.

17. Zhang Z, Zhang A, Li M, Zhao H, Karypis G, Smola A. Multimodal chain-of-thought reasoning in language models. Trans Mach Learn Res 2024.

18. Wang X, Wei J, Schuurmans D, et al. Self-consistency improves chain of thought reasoning in language models. In: Proceedings of the International Conference on Learning Representations (ICLR), 2023.

## Supplementary Figures

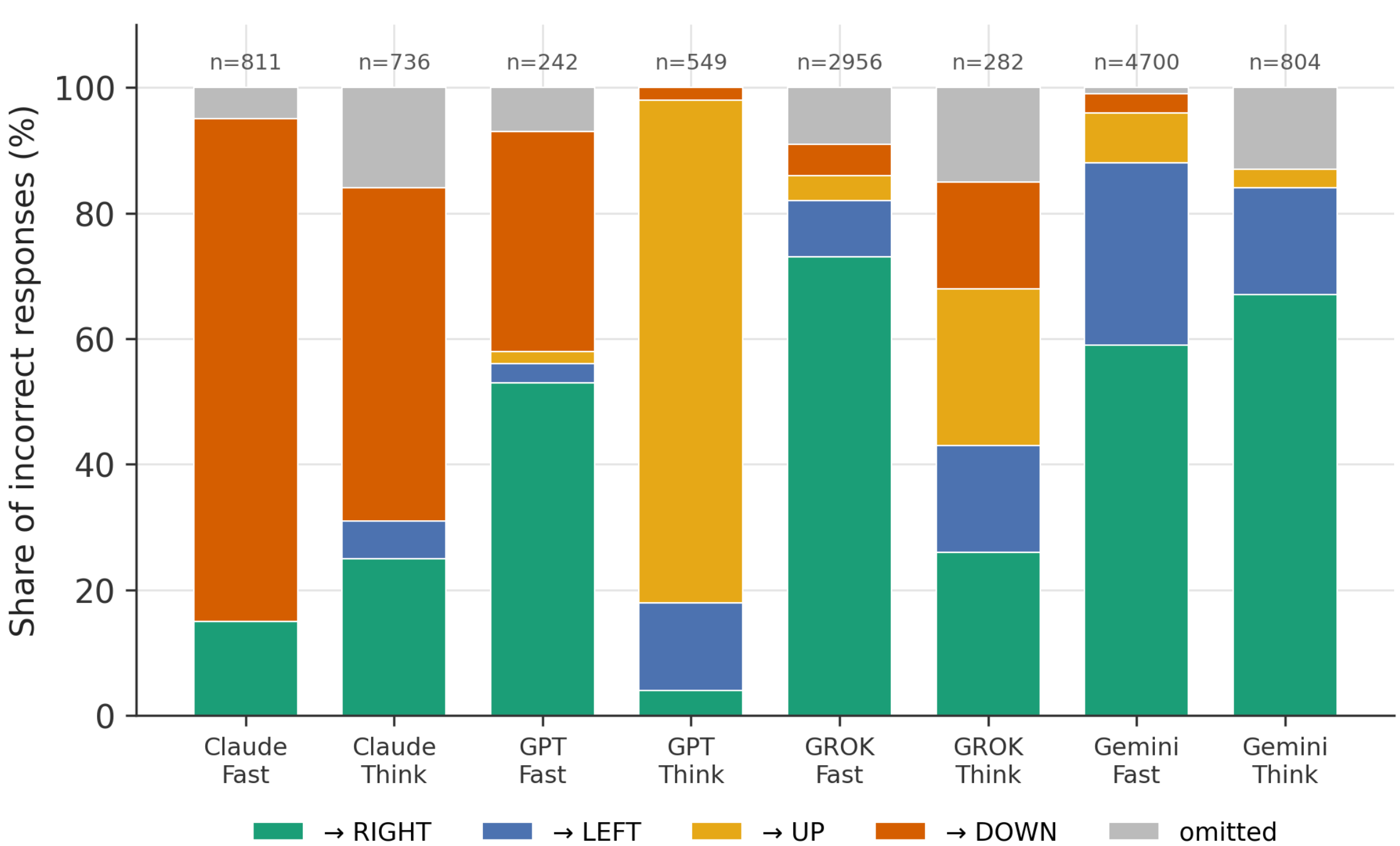

***Figure S1.*** *Directional structure of glyph-level errors by model and reasoning mode. Each 100% stacked bar shows the share of incorrect responses assigned to RIGHT, LEFT, UP, DOWN, or omitted; n above each bar is the total number of incorrect responses in that cell.*

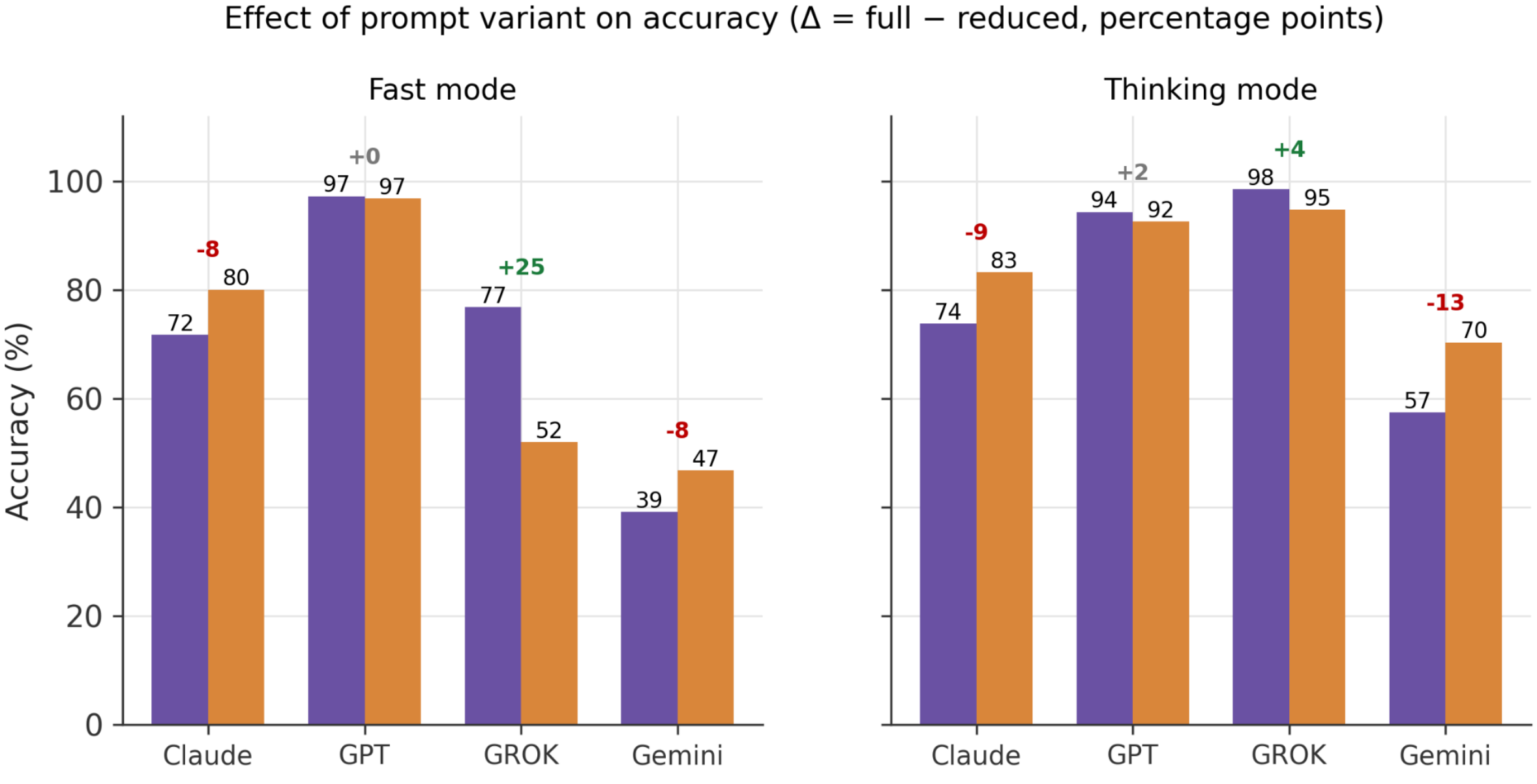

***Figure S2.*** *Effect of prompt variant on glyph-level accuracy. Bars compare the full structured prompt (rule present) with the reduced prompt (rule absent) separately in Fast and Thinking modes. Delta labels report full minus reduced accuracy in percentage points.*

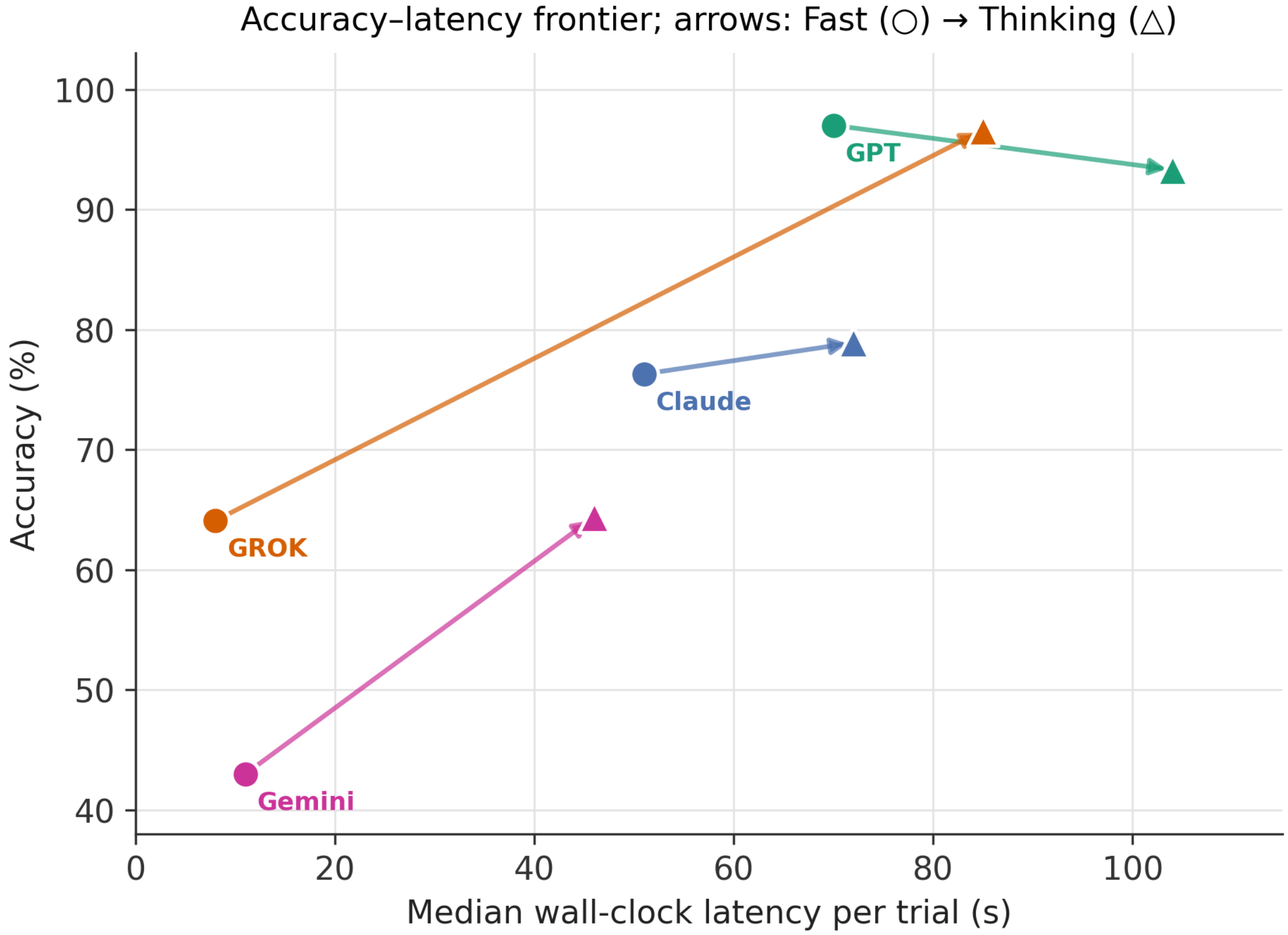

***Figure S3.*** *Accuracy-latency frontier across reasoning modes. Circles show Fast mode and triangles show Thinking mode; arrows connect each model's Fast and Thinking operating points. Latency is the median wall-clock time per trial.*

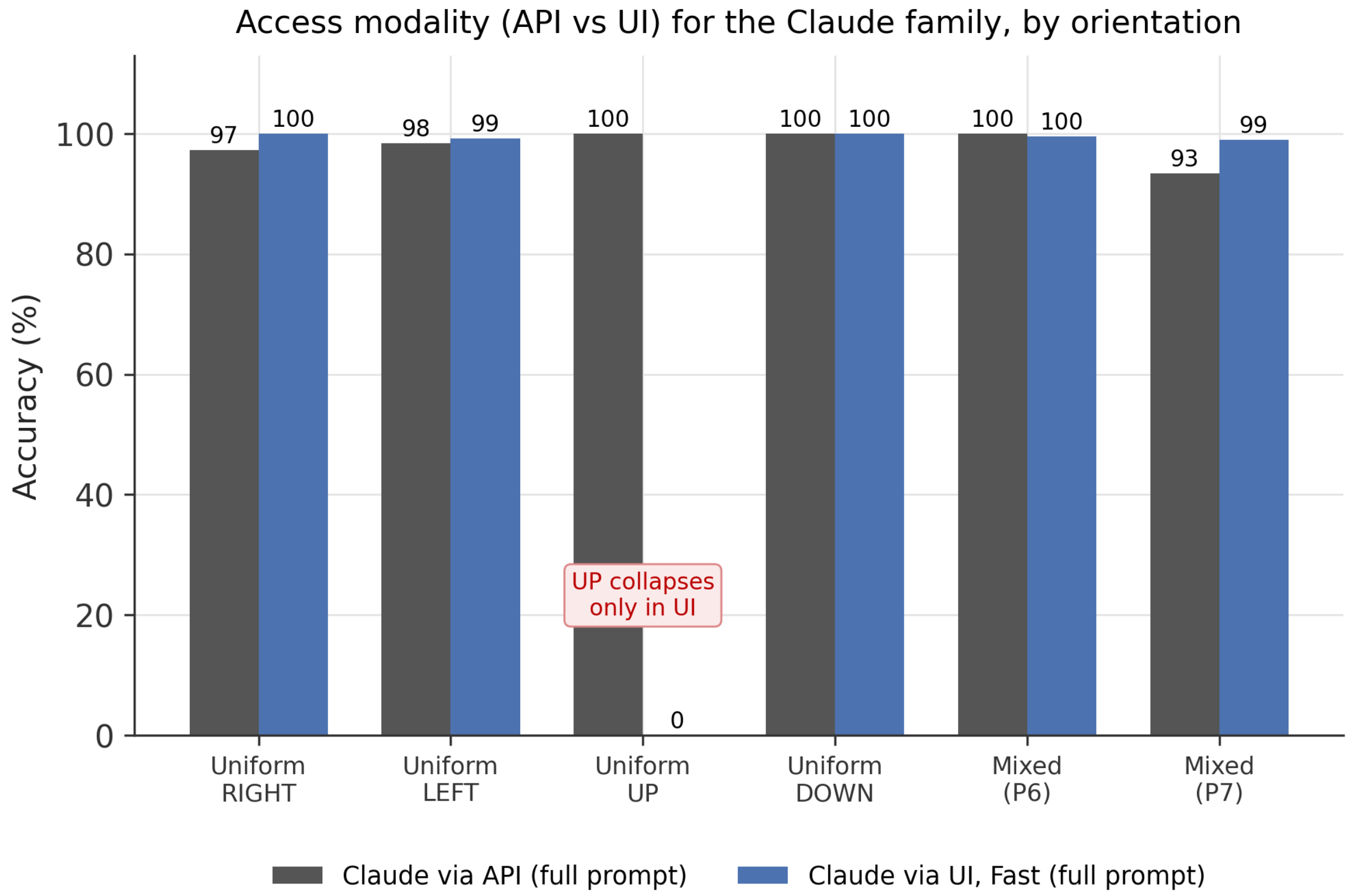

***Figure S4.*** *Access-modality comparison for the Claude family by stimulus orientation. Bars compare Claude via API under the full prompt with Claude via the consumer UI in Fast mode under the full prompt. The API-UI gap is concentrated on the uniform-UP chart.*